\documentstyle[12pt]{article}
\title{A consistent electromagnetic duality}
\author{Carlos A. P. Galv\~{a}o  \thanks{\noindent e-mail :
         carlosap@fis.unb.br}  \\
        Depto. de F\'{i}sica, U. F. do Rio Grande do Norte  \\ 
        CEP 59075-970  Natal - RN , Brazil \footnote{On leave from
           Depto. de F\'{i}sica, Univ. de Bras\'{i}lia} \\
        Juan A. Mignaco \thanks{\noindent e-mail:
                mignaco@if.ufrj.br  }   \\
        Inst. de F\'{i}sica, Univ. Fed. do Rio de Janeiro     \\
        CEP 21945-970 Rio de Janeiro - RJ  , Brazil  }
\date{}
\begin{document}
\maketitle
\begin{abstract}

We present a new view for duality in classical electromagnetic theory, 
based on the physical properties of a dual theory, eliminating the problems of the usual treatment of the subject.

\end{abstract}

\vspace{1.0cm}


\vspace{1.0in}

The concept of duality has received considerable attention in gauge theories. It provides 
useful tools to construct solutions to the field equations, namely, those which are self-dual or anti-self-dual, or allows to study regimes of the theory which prevent the use of perturbation expansions \cite{rschz}. 

Duality in classical electromagnetic theory was discovered by Heaviside \cite{rhea}
a century ago for the Maxwell equations in vacuum. He saw that 
\begin{equation}
     \left.
      \begin{array}{rl}
         \nabla \wedge {\bf E}  =  &  -  {\partial {\bf B} \over \partial t}  \\
                                   &                                          \\
         \nabla \wedge {\bf B}  =  &  \frac{1}{c^{2}} {\partial {\bf E} \over 
                                            \partial t} 
      \end{array}
     \right\}     \,  .
\label{eq;max0}
\end{equation}
exchanged among themselves under the replacements
\begin{equation}
      { \bf E}  \rightarrow  - c {\bf B}  \,  ,
\hspace{1.0in}
       c {\bf B}  \rightarrow    {\bf E}  \,  .
\label{eq;dual}
\end{equation}

This symmetry of the system, duality (we shall refer to it as Heaviside duality) ,
originated a lot of speculation about its meaning: is the electric field equivalent to the magnetic induction field, and the reverse?. With the advent of nonabelian gauge theories for elementary 
particle physics, a lot of work in Physics and Mathematics has been performed to clarify the meaning and applicability of duality, as exposed above, or in its modern nonabelian versions. 

Let us recall that the original fields in the Maxwell equations have electric charges or currents and/or time varying fields as their sources. Even in vacuum, electric fields are considered the ones accelerating electric charges parallel to their direction, whereas magnetic fields provide transverse acceleration for electric charges. 

Magnetic materials are related to elementary magnetic dipoles but single isolated magnetic charges have never been observed. Dirac \cite{dir} introduced magnetic monopoles in this framework to provide sources for the magnetic induction field, i. e.,
\begin{equation}
   \nabla \cdot {\bf B} = g {\mu}_{0} \delta ( {\bf x} )
\end{equation}

To preserve the relation with the magnetic potential,
\begin{equation}
     {\bf B} = \nabla \wedge \ {\bf A}
\end{equation}
a topological structure had to be included (\cite{greu}, \cite{yang}). It determined the famous Dirac relation between the electric charge of a test particle and the strength of the monopole field :
\begin{equation}
  q g = 2 \pi {\it n} \frac{\hbar}{{\mu}_{0}}    \,  .
\end{equation}
It is usually assumed that the introduction of magnetic monopoles is related to Heaviside duality, though the relation is somewhat vague 
\footnote{It is worth pointing that the same quantum of flux appears here as in the proposed Aharonov-Bohm \cite{aha} effect}
.

There remains unwanted aspects of the theory at the electromagnetic level. Standing high is the fact that the lagrangian of the theory changes sign under duality, that is, Heaviside duality is a symmetry for the equations of motion but not for the lagrangian providing them. Neglecting sources,
\begin{equation} 
     L \{ {\bf E}, {\bf B} \}  =
      \frac{\epsilon_{0}}{2} \int d^{3}x \left( {\bf E}^{2} - c^{2}{\bf B}^{2} \right)
\end{equation}

Heaviside duality may be extended to a continuous variation \cite{larmor}. The generator for infinitesimal transformation has been proposed to be non-local \cite{destei} or built breaking explicit Lorentz invariance \cite{schsen} 

In this article we present a new interpretation for duality which provides a consistent physical 
picture and avoids all the problems alluded before. The matter being largely speculative, our point is, notwithstanding, the physical basis for the symmetry.

Let us look at the equations for would be classical magnetodynamics, the dual theory to real electrodynamics. They are \cite{schw}:
\begin{equation}
  \left.
     \begin{array}{rl}
   \nabla \cdot {{\bf E}^{\prime}}  =  & 0                                  \\
                                       &                                    \\
   \nabla \cdot {{\bf B}^{\prime}}  =  & \mu_{0} \rho_{m}                   \\
                                       &                                    \\
   \nabla \wedge {{\bf E}^{\prime}}  = & - \frac{1}{\epsilon_{0}} {\bf j}_{m}
                   -  {\partial {{\bf B}^{\prime}} \over \partial t}        \\
                                       &                                    \\
   \nabla \wedge {{\bf B}^{\prime}}  = & \frac{1}{c^{2}} {\partial {{\bf E}^{\prime}}
                                      \over {\partial t}}
     \end{array}
        \right\}   \;  .
\end{equation}

The primes indicate that though the symbols appear in their positions for the original 
Maxwell electrodynamics equations, they satisfy different equations than in electrodynamics.
The physical content of the equations is different: the sources for the magnetic field are now magnetic charges and/or currents, and time variations of the electric field. The latter, now, is generated only by magnetic currents or by the time variation of the magnetic field.

With the same arguments of the usual case, an axial vector electric potential and a pseudo-scalar magnetic potential may be introduced \cite{cabfer} from the homogeneous equations:
\begin{equation}
  \left.
   \begin{array}{rl}
        {{\bf E}^{\prime}} =  &  \nabla \wedge {\bf W}                     \\
                              &                                            \\
        {{\bf B}^{\prime}} =  &  -  {{\partial \psi} \over {\partial t}}  +
                 \frac{1}{c^{2}} {\partial {\bf W} \over {\partial t}}  \\
                           &
   \end{array}
  \right\}  \,  .
\end{equation}

With these potentials, a second order lagrangian may be constructed from which the Euler-Lagrange equations are the ones above for the fields and their magnetic charges and currents. In three dimensional notation, the lagrangian is:
\begin{equation}
   L \{ {{\bf B}^{\prime}}, {{\bf E}^{\prime}} \}  =  \frac{\epsilon_{0}}{2} \int d^{3}x 
         \left( c^{2} {{\bf B}^{\prime}}^{2} - {{\bf E}^{\prime}}^{2} \right)  - 
         \rho_{m} \, \psi  -  {\bf j}_{m} \cdot {\bf W}
\end{equation}

One can easily obtain in covariant four-dimensional notation the corresponding lagrangian, envolving an anti-symmetric tensor of second rank, $G_{\mu \nu}$. Its space-space components are the components of the electric (primed) field, while the time-space ones are  
\begin{equation}
                 G_{0k} = c {B^{\prime}}_{k}
\end{equation}

This exchange of components with respect to the corresponding tensor in electrodynamics, 
$F_{\mu \nu}$, makes tempting to identify one as the dual of the other. This is not allowed, the corresponding fields satisfy different differential equations and have correspondingly different physical content. It may be correct to say that in the magnetodynamic case the lagrangian density is written in terms of the dual of what would be the field intensity tensor with the {\underline {same}} fields.

We are now in position to analyze duality. The wave equations for the fields in electrodynamics and magnetodynamics are the same. That is, in this harmonic sector (in the sense of differential forms) the meaning of the exchange in the original Heaviside duality has the meaning of an exchange between the magnetodynamic and electrodynamic equations:
\begin{equation}
   {\bf E} \rightarrow - c {{\bf B}^{\prime}}
\hspace{1.0in}
   c {\bf B} \rightarrow {{\bf E}^{\prime}}
\end{equation}
and the corresponding ones for the other fields. 

One can write a lagrangian density with satisfactory dual symmetry under the exchange of electric and magnetic fields in the electrodynamic and the magnetodynamic sectors. It is
\begin{equation}
   {\cal L} =  - \frac{\epsilon_{0}}{4} F_{\mu \nu} F^{\mu \nu} 
               - \frac{\epsilon_{0}}{4} G_{\mu \nu} G^{\mu \nu}  
            - \frac{1}{c} {j_{e}}_{\mu} A^{\mu}  - \frac{1}{c} {j_{m}}_{\mu} W^{\mu} 
            +  (gauge fixing terms)  \, .
\end{equation}

Our extended version for duality exchanges the terms in the lagrangian. Of course, for the 
free space equations, one has Heaviside duality in both sectors.
One can also add monopoles by looking for topologically non-trivial configurations for the potentials. Monopoles are seen as not directly related to the extended electromagnetic duality.

The generator of duality continuous transformations may be easily obtained. From the lagrangian,
the corresponding momenta to the potentials $A^{\mu}$ and $W^{\mu}$ are
\begin{equation}
 \left.
  \begin{array}{rlrl}
       \Pi_{\mu} =  &  \frac{\partial {\cal L}}{\partial (\partial^{0} A^{\mu})} = & 
                   - \epsilon_{0} E_{k} \delta_{k \mu}                                     \\
    {\Pi^{\prime}}_{\mu} = &  \frac{\partial {\cal L}}{\partial (\partial^{0} W^{\mu})} = &
                   - \epsilon_{0} c {B^{\prime}}_{k} \delta_{k \mu}
       \end{array}
    \right\}  \,  .
\end{equation}
In terms of the fields and their momenta, the transformations, parametrized by the continuous 
transformations as
\begin{equation}
  \left.
     \begin{array}{rl}
        {\bf E} = & \cos \eta {\bf E} - \sin \eta c {{\bf B}^{\prime}}          \\
                   &                                                         \\
       c {\bf B} = &  \sin \eta {{\bf E}^{\prime}} + \cos \eta c {\bf B}
    \end{array}
  \right\}
\end{equation}
and the corresponding ones for the other couple of fields, are obtained from the generator:
\begin{equation}
           M =  \int d^{3}x [ A^{\mu} {\Pi^{\prime}}_{\mu} 
                          -   \Pi_{\mu} W^{\mu} ]
\end{equation}
which is perfectly local and covariantly written.

In conclusion, to make a consistent mathematical treatment of electromagnetic duality, preserving locality, Lorentz invariance and positivity of the energy,the physical properties of dual systems must be enlarged correctly. 

Our proposal looks very similar to what is known as Hodge lemma in mathematical terms \cite{hodge}: the space of second rank tensors in a compact manifold (we work in non-compact four-dimensional spacetime) decomposes in harmonic terms 
(solutions to the wave equation), curl of vectors ({\bf A}, in our case) and divergences of higher rank tensors (or their duals, in our case, {\bf W}). The spaces of this decomposition are mutually orthogonal.

It is true that Nature presents us only electric charges and currents, but duality points for a complementary world closely related to the former. The fact that this symmetry is not apparent in the real world points to some kind of breaking, and may be a fruitful field of research.

\vspace{1.0in}
{\it Acknowledgments} We acknowledge partial support from CNPq during part of this work, and discussions with Prof. C. J. Wotzasek.

\end{document}